\documentclass[11pt]{article}

\usepackage[letterpaper,margin=1in,headheight=16pt]{geometry}
\usepackage[T1]{fontenc}
\usepackage{lmodern}
\usepackage{graphicx}
\usepackage{booktabs}
\usepackage{enumitem}
\usepackage{titlesec}
\usepackage{fancyhdr}
\usepackage{amsmath}
\usepackage{amssymb}
\usepackage[numbers,sort&compress]{natbib}

\newcommand{\doi}[1]{doi: \href{https://doi.org/#1}{#1}}
\newcommand{\napo}{N\={a}p\={o}waw\={a}\textquoteleft enakaulua}

\newcommand{\ergs}{\ifmmode \textrm{erg\ s}^{-1} \else erg s$^{-1}$\fi}
\newcommand{\Msun}{\ifmmode \textrm{M}_{\odot} \else $M_{\odot}$\fi}
\newcommand{\Lsun}{\ifmmode \textrm{L}_{\odot} \else $L_{\odot}$\fi}

\usepackage{xcolor}
\definecolor{genesisgold}{HTML}{B5771A}
\definecolor{genesisink}{HTML}{10131F}
\definecolor{genesisgray}{HTML}{6B7386}

\usepackage[colorlinks=true,linkcolor=genesisgold,citecolor=genesisgold,urlcolor=genesisgold]{hyperref}

\titleformat{\section}{\normalfont\large\bfseries\color{genesisink}}{}{0pt}{}
\titlespacing*{\section}{0pt}{1.3em}{0.45em}

\fancypagestyle{genesisfirst}{%
  \fancyhf{}%
  \fancyhead[L]{}%
  \fancyhead[R]{\small\textcolor{genesisgray}{MCT Concept White Paper}}%
  \fancyfoot[C]{\small\thepage}%
}

\begin{document}
\thispagestyle{genesisfirst}

{\LARGE\bfseries\color{genesisink} 
Dual and Binary AGN at Cosmic Dawn:  The need for a JWST  Program to resolve the merger channel of Early Black Hole Growth
\par}
\vspace{0.8em}

{\large 
Filippo Mannucci\textsuperscript{1}, 
Matteo Ceci\textsuperscript{2,1},
Norman A. Grogin\textsuperscript{3},
Anton M. Koekemoer\textsuperscript{3},
Alessandro Marconi\textsuperscript{2,1},
Enik\H o Reg\H os\textsuperscript{4},
Raffaella Schneider\textsuperscript{5},
Martina Scialpi\textsuperscript{1},
Giulia Tozzi\textsuperscript{6},
Alessandro Trinca \textsuperscript{7,3} 
Giacomo Venturi\textsuperscript{1},
Maria~Vittoria Zanchettin\textsuperscript{8},
Huanian Zhang\textsuperscript{9},
Muhammad A. Latif\textsuperscript{10}
 \par}
\vspace{0.5em}

{\small\itshape\textcolor{genesisgray}{%
\textsuperscript{1}INAF - Osservatorio Astrofisico di Arcetri - Italy \quad
\textsuperscript{2} Dipartimento di Fisica e Astronomia, Università di Firenze, Italy \quad
\textsuperscript{3}Space Telescope Science Institute \quad
\textsuperscript{4}Konkoly Observatory HUN-REN \quad
\textsuperscript{5}Dipartimento di Fisica, Sapienza Università di Roma \quad
\textsuperscript{6} Max-Planck-Institut für extraterrestrische Physik, Germany \quad
\textsuperscript{7} INAF - Osservatorio Astronomico di Roma, Italy \quad
\textsuperscript{8} INAF - Istituto di Astrofisica e Planetologia Spaziali - Italy \quad
\textsuperscript{9} Department of Astronomy, School of Physics, Huazhong University of Science and Technology, Luoyu Road 1037, Wuhan, China \quad
\textsuperscript{10} Physics Department, College of Science, United Arab Emirates University, PO Box 15551, Al-Ain, UAE \quad 
\vspace{0.4em}
}}

{\small Corresponding author: Filippo Mannucci (\href{mailto:afilippo.mannucci@inaf.it}{filippo.mannucci@inaf.it})\par}

\vspace{1.0em}

\noindent\textbf{\color{genesisink}Abstract.}\enspace

The discovery with JWST of accreting black holes (BHs) at $z=4$--$9$ in numbers far exceeding pre-JWST expectations, many of them overmassive relative to their hosts and linked to the enigmatic Little Red Dots (LRDs), raises the question of how early BHs form and grow. Hierarchical assembly predicts that galaxy mergers drive gas to galactic nuclei and bring BHs together, leaving a direct observational signature: dual AGN at tens of kpc separations and sub-kpc to few-kpc ``duets'' in post-merger galaxies, which are the progenitors of the low-frequency GW sources targeted by pulsar timing arrays and LISA. JWST has opened the window to study these systems at $z>3$.  However, these early discoveries are very limited number and come from heterogeneous, largely serendipitous samples, so these systems can hardly be used for robust  statistical studies. Within the {\it Genesis} initiative We propose a uniformly selected census of dual AGN and their hosts from cosmic noon to cosmic dawn, as a science case for the {\it Genesis} initiative. Our tiered strategy uses the survey imaging and spectroscopy for candidate selection at no extra cost, followed by NIRSpec IFU confirmation, multi-band NIRCam (and MIRI/ALMA) host characterisation, and multi-epoch revisits. Confirming and characterising $\sim\!100$ pairs requires of order $10^3$\,hr, well beyond the ceiling of standard TAC programs, and demands multi-cycle continuity. The census will deliver pair fractions as functions of redshift, separation and luminosity, BH masses and mass ratios, host properties and environments, and the empirical anchor needed to turn merger models into realistic LISA rate forecasts.

\section{Scientific Justification}

Among the most debated questions in modern astrophysics \cite{genesis26} 
are when and how the first massive black holes (BHs) form and grow, how they are related to the enigmatic Little Red Dots (LRDs), and when the scaling relations between BHs and their host galaxies first come into place. JWST has shown that accreting BHs are far more common at early times than expected: AGN at $z=4$--9 exceed the extrapolated bolometric and UV luminosity functions of pre-JWST campaigns by factors of 10--100~\cite{caceresburgos}, and many appear overmassive relative to their hosts when compared to local scaling relations~\cite{pacucci}. Current explanations invoke heavy seeds ($10^3$--$10^5\,\Msun$) growing at the Eddington limit, or light seeds undergoing super-Eddington episodes~\cite{caceresburgos,volonteri25}. In either case, hierarchical assembly predicts that galaxy mergers channel gas towards galactic nuclei, feed nuclear accretion and bring BHs together~\cite{dimatteo,volonteri22}. This is the \emph{merger channel} of BH growth.

The merger channel leaves a direct observational signature: pairs of accreting BHs. At tens of kpc separations, when still in pre-merging galaxies, they appear as dual AGN. 
At 1--10 kpc separations, 
when they are part of the same post-merging galaxy and where dynamical friction in dense, compact systems is efficient, they are dubbed `duets'. These objects are on their way to coalescence with the emission of the low-frequency gravitational-wave (GW) signals targeted by pulsar timing arrays~\cite{agazie} and LISA~\cite{amaroseoane,colpi}. Dual AGN and duets are therefore the most direct electromagnetic census of the progenitors of BH mergers, and at the same time a test of BH--host co-evolution.

In this white paper, we argue that a systematic, uniformly selected census of dual AGN and their host galaxies, from cosmic noon to cosmic dawn, is a high-impact science case for the {\it Genesis} initiative \cite{genesis26}. We further argue that it cannot be achieved with the heterogeneous, largely serendipitous samples that the standard TAC process has produced so far.

\section*{Dual AGN and Duets with JWST: a Window Just Opened}
At cosmic noon, Gaia has allowed for systematic searches for sub-arcsec dual AGN  \cite[e.g.,][]{mannucci22,gross25,scialpi26a,zanchettin26} and the first systematic comparisons between observations and model predictions  \cite[e.g.,][]{scialpi26b,puertosanchez25}.
At $z>3$ known quasar pairs are rare, JWST has pushed the frontier to $z>4$ and to sub-kpc separations (Table~\ref{tab:sys}). Five results define the current state of the art.

\textbf{Dual AGN are not as rare as expected.} Among 16 AGN observed with NIRSpec IFU at $2<z<6$, 3--5 are multiple-AGN systems with separations of 3--28 kpc, a fraction that moderately exceeds the $\sim$10\% predicted by cosmological simulations~\cite{perna25}. In the overdense field of BR1202-0725 at $z=4.7$, a QSO and a submillimetre galaxy hosting an AGN are separated by 24 kpc, and at least ten galaxies lie within 980 kpc$^2$~\cite{zamora}. A dual AGN candidate at $z\sim5.4$ (10.4 pkpc) sits in a Ly$\alpha$ nebula at the center of an overdensity, and the inferred dual fraction is higher than theoretical expectations~\cite{li25}.

\textbf{Sub-kpc duets exist at $z>4$.} NIRSpec IFU channel maps isolated two possible broad-line regions (BLRs) separated by $190\pm40$ pc, plus a third one at 1.7 kpc, in a $z=5.02$ galaxy. The three BH masses are $\log M_\bullet/\Msun=7.9$, 5.8 and 6.3, and a simple dynamical-friction estimate gives a merger time $\lesssim700$ Myr~\cite{ubler25}. Other IFU pairs are known at $z=7.2$ (separation $\sim$620 pc)~\cite{ubler24} and $z=4.5$ ($\sim$1.5 kpc)~\cite{suh26}. Asymmetric broad-line profiles in integrated MSA and slitless spectra are consistent with further sub-kpc pairs~\cite{maiolino24,matthee24}, but only IFU data can separate the BLRs.

\textbf{LRDs can come in pairs.} Several instances of Dual-LRD candidates with small separations have been found 
\cite{tanaka24,merida25,yanagisawa26}. 
Two have slitless redshifts $z=5.82$ and 5.46 (1.6 and 7.4 kpc), and the sub-arcsecond angular correlation function exceeds the extrapolation from 10--100$''$ scales by a factor $\sim$20--30~\cite{tanaka24}.

\textbf{Even luminous quasars pair up.} Visual inspection of the NIRSpec IFU cubes of $\sim$200 quasars at $5.7<z<7$ revealed the pair \napo{} ($z=6.08$, 0.98$''$ = 5.56 pkpc, $\Delta v\simeq15$ km s$^{-1}$, $M_\bullet\simeq1.5\times10^9$ and $4\times10^8\,\Msun$, $\lambda_{\rm Edd}\simeq1.2$ and 1.1). Its occurrence exceeds extrapolated high-$z$ clustering by 2--3 orders of magnitude~\cite{onorato}.

\textbf{Duet hosts at cosmic noon are surprising.} The 3.8 kpc dual quasar J0749+2255 at $z=2.17$ shows ionised-gas kinematics of a rotating disk rather than a disturbed merger~\cite{ishikawa}. Its PAH-based star formation rate is $\sim10^{3}\,\Msun$ yr$^{-1}$, yet its molecular gas content falls short of the Kennicutt--Schmidt expectation~\cite{chenYC24}. A $z\sim2$ dual-AGN host has been reported to be bulge-less~\cite{chenYC23}, whereas simulations predict hosts of dual AGN to lie below the BH--stellar mass relation~\cite{volonteri22,chenN23}.

\begin{table}[t]
\centering\small
\renewcommand{\arraystretch}{1.12}
\begin{tabular}{@{}p{4.1cm}p{1.5cm}p{3.0cm}p{3.4cm}l@{}}
\toprule
System & $z$ & Proj. separ. & JWST data & Ref. \\
\midrule
GA-NIFS multiple AGN  & 2--6 & 3--28 kpc & NIRSpec IFU & \cite{perna25}\\
J0749+2255 (dual quasar) & 2.17 & 3.8 kpc & NIRSpec+MIRI IFU & \cite{ishikawa,chenYC24}\\
BR1202-0725 & 4.7 & 24 kpc & NIRSpec IFU & \cite{zamora}\\
Dual AGN in Ly$\alpha$ nebula & $\sim$5.4 & 1.7$''$ (10.4 kpc) & NIRCam, NIRSpec & \cite{li25}\\
Dual-LRD candidates & 5.46, 5.82 & 7.4, 1.6 kpc & NIRCam, slitless & \cite{tanaka24}\\
BH pair (single galaxy) & 4.5 & $\sim$1.5 kpc & NIRSpec IFU & \cite{suh26}\\
Triple BH, J0148$-$4214 & 5.02 & 0.19 and 1.7 kpc & NIRSpec IFU & \cite{ubler25}\\
BH pair & 7.2 & $\sim$0.62 kpc & NIRSpec IFU & \cite{ubler24}\\
\napo{}  & 6.08 & 5.56 kpc & NIRSpec IFU & \cite{onorato}\\
\bottomrule
\end{tabular}
\caption{Examples of dual and multiple AGN systems at $z>2$ discovered with JWST. Systems were found in very different ways (serendipity, parent-sample IFU surveys, imaging morphology, slitless spectra, Gaia).}
\label{tab:sys}
\end{table}

\section*{What the Current Data Cannot Tell Us}
The discoveries above are exciting, but they cannot yet answer the questions that matter, for five reasons.
\begin{itemize}
\item \textbf{No comparable statistics.} Quoted multiplicity fractions range from $\sim$0.5\% (one pair among $\sim$200 luminous quasars, a lower limit) to 20--30\% (3--5 of 16 AGN)~\cite{onorato,perna25}. These numbers differ in luminosity, separation, redshift and selection, so they cannot be compared, nor confronted with simulations, until a uniformly selected sample exists.
\item \textbf{Incomplete sub-kpc coverage.} The NIRSpec IFU PSF ($\approx0.15''$, $\approx$0.8 pkpc at $z\sim6$) leaves sub-kpc pairs undetected in survey-mode searches~\cite{onorato}. The $z=5$ pair candidate at 190 pc was separated only through spectral channel maps of the BLR~\cite{ubler25}, and sub-kpc candidates identified in 1D spectra remain unconfirmed.
\item \textbf{Unknown host galaxies.} For most pairs we do not know whether the two BHs share one host or two, nor their stellar masses, morphologies or gas fractions. Measuring them requires removing two bright point sources from multi-band imaging with a stable, well-modelled PSF.
\item \textbf{Pairs versus lenses and chance projections.} Differential microlensing can mimic spectral differences between components, and neither J0749+2255 nor \napo{} can be firmly classified as physical pairs from spectra alone~\cite{ishikawa,onorato}. Statistical samples with narrow-line kinematics are the way forward.
\item \textbf{No empirical anchor for theory.} Predicted BH merger rates at $z\ge5$ usually neglect sub-kpc dynamics and are explicitly optimistic~\cite{caceresburgos}. The first systems already span mass ratios from $\approx$4 to $\approx$100~\cite{onorato,ubler25}. Observed separations, mass ratios and velocity offsets are the missing input to turn such models into realistic LISA rates.
\item \textbf{Large-scale environments.} Due to the very limited number of dual AGNs, estimating the large-scale environments of these objects through cross-correlation with comparable galaxies at the same cosmic epoch \cite{ZhangC2026}, and thereby constraining their host dark matter halos, is almost impossible.
\end{itemize}

\section*{Why Only a Multi-cycle Treasury Scale Will Do}
\textbf{Uniform parent samples.} The {\it Genesis} program will, by design, include very deep imaging and an extremely large spectroscopic sample with uniform selection. This is exactly what the dual-AGN census requires. Pair fractions as a function of redshift, separation and luminosity are meaningful only when completeness is quantified, and simulations can be compared with observations only when selection is uniform. The same applies to dual LRDs, whose sub-arcsecond clustering excess~\cite{tanaka24} is a new constraint on how LRDs are triggered, complementary to the debate on their nature~\cite{begelman,naidu26,chisholm}.

\textbf{Time beyond the TAC ceiling.} The largest extragalactic GO programs so far (NEXUS, 368 hr; VENUS, 303 hr; COSMOS-Web, 255 hr) mark the practical ceiling of what TACs deliver. Dual-AGN science needs more than this because the optimal strategy has several tiers, each with a real cost. Bright quasars need $\sim$45 min of NIRSpec IFU time~\cite{onorato}. Faint broad-line AGN pairs at $z\sim4$--5 need several hours of R$\sim$2700 IFU exposure each, of order 3--9 hr of science time (5--11 hr charged), to separate BLRs at few 100s pc. Four-band NIRCam imaging of the hosts of $z\sim1.3$--2.4 pairs costs about 1 hr of science time (2.8 hr charged) per target. Confirming and characterising $\sim$100 pairs would therefore take of order $10^3$ hr (roughly 500--1100 hr of IFU time plus $\sim$300 hr of NIRCam time). This is an illustrative scaling, not a design, but it shows why the census must be planned as part of a multi-cycle program rather than as stand-alone GO proposals.

\textbf{Multi-cycle continuity and time domain.} Candidates are identified in the survey data and must then be followed up, so the work is sequential by nature. Multi-epoch revisits, which probe BLR variability and test lensing hypotheses, require cadenced observations that {\it Genesis} itself identifies as structurally impossible within a single cycle.

\textbf{Time-critical synergies.} MIRI is already at reduced efficiency at long wavelengths, and MIRI spectroscopy has not yet been deployed over large samples, yet it provides the star-formation tracers at cosmic noon~\cite{chenYC24}. ALMA provides the cold-gas masses needed to determine the gas fractions of dual-AGN hosts. A program in the next few cycles will also provide targets for ELT and GMT and the LISA era.

\section{Multi-tier strategy}

We sketch a tiered strategy that makes the census affordable by allowing the {\it Genesis} survey data \cite{genesis26} to do most of the selection.
\begin{itemize}
\item \textbf{Tier 1: selection within {\it Genesis} data (no additional time).} Search the imaging for multi-component compact sources using pixel-by-pixel color selection~\cite{tanaka24}. Flag asymmetric or double-peaked broad lines and spatially offset narrow lines in the MSA and grism spectra~\cite{maiolino24}.
\item \textbf{Tier 2: IFU confirmation.} NIRSpec IFU at R$\sim$2700 for candidate pairs at $z\gtrsim4$. Targets should be chosen by a documented selection function, so that confirmed and refuted candidates both feed the completeness calibration.
\item \textbf{Tier 3: host characterization.} Multi-band NIRCam imaging with PSF libraries built from field stars, and MIRI where star formation or obscuration must be traced, combined with ALMA gas masses.
\item \textbf{Tier 4: multi-epoch revisits} of the confirmed systems, over several years in the observed frame.
\end{itemize}

\section*{Science Returns}
This architecture would deliver: (i) dual and multiple AGN fractions as a function of $z$, separation and luminosity, tested against cosmological simulations; (ii) BH masses, mass ratios and velocity offsets of resolved pairs, constraining dynamical-friction timescales and the input to LISA GW event rate forecasts; (iii) whether pairs share the same host galaxy, plus BH--stellar mass relations and gas fractions of dual-AGN hosts, including mass decompositions that correct masses derived from unresolved profiles; (iv) the environments of pairs, building on cases already found in overdense regions~\cite{zamora,li25}; and (v) the small-scale clustering of quasars and LRDs~\cite{onorato,tanaka24}.



\end{document}